\documentstyle[aps,prl,floats,twocolumn,epsf]{revtex}

\begin{document}
\draft

\twocolumn[
\hsize\textwidth\columnwidth\hsize\csname
@twocolumnfalse\endcsname

\title{Quantum Dynamics of Pseudospin Solitons in
       Double-Layer Quantum Hall Systems}

\author{Jordan Kyriakidis$^1$, Daniel Loss$^1$, and
        A.~H. MacDonald$^2$}

\address{$^1$Department of Physics and Astronomy,
             University of Basel, Klingelbergstrasse 82,
             CH--4056 Basel, Switzerland\\
         $^2$Department of Physics, Indiana University,
             Bloomington, IN 47405, U. S. A.}

\date{\today}
\maketitle

\begin{abstract}
  Pseudospin solitons in double-layer quantum Hall systems can be
  introduced by a magnetic field component coplanar with the electrons
  and can be pinned by applying voltages to external gates. We
  estimate the temperature below which depinning occurs predominantly
  via tunneling and calculate low-temperature depinning rates for
  realistic geometries.  We discuss the local changes in charge and
  current densities and in spectral functions that can be used to
  detect solitons and observe their temporal evolution.
\end{abstract}

]

\narrowtext

The study of multicomponent Quantum Hall systems~\cite{halperin83} has
been enriched by the discovery of a variety of new phases. In double
layers, the relevant discrete degrees of freedom are labelled by the
electron's layer and spin indices. At Landau level filling factor
$\nu=2$, recent theoretical work predicted several interesting
phases~\cite{zheng97} in which both layer and spin play a role; the
existence of these phases has been confirmed by
experiment~\cite{pellegrini97}. The present study is on double-layer
systems at filling-factor $\nu =
1$~\cite{macdonald90a,doublelayerrefs}.  At this filling factor, the
low-energy electron states are spin-polarized and the system has a
broken symmetry ground state with spontaneous interlayer phase
coherence~\cite{murphy94}.  The rich phase diagram for these systems,
including the effects of in-plane fields, has been discussed at length
in Ref.~\cite{doublelayerrefs}.

It is useful to describe this system using a pseudospin
language~\cite{macdonald90a,doublelayerrefs} in which pseudospin-up
(-down) refers to an electron in the top (bottom) layer.  The action
is that of a two-dimensional ferromagnet with a hard-axis anisotropy
and a Zeeman field perpendicular to the
hard-axis~\cite{doublelayerrefs}. The pseudospin configuration is
specified by the spherical-coordinate fields $\theta (x,y)$, which
describes the difference in charge density between the layers, and
$\phi (x,y)$, which describes the relative phase of electrons in top
and bottom layers. Phase solitons $\phi_0(x)$ exist as solutions to
the classical equations of motion~\cite{doublelayerrefs}.  In this
Letter we address the quantum dynamics of such solitons, predicting
that, when pinned by applying gate voltages, depinning occurs at
accessible temperatures predominantly via quantum tunneling.  This
system offers a number of advantages for macroscopic quantum tunneling
studies, especially the possibility of using gate voltages and
in-plane fields in combination to control metastable-state placement.
We also discuss several local properties which can be used to detect
solitons and observe their temporal evolution.

Neglecting for the moment the effect of in-plane fields, the leading
contributions to the imaginary-time effective action for the
pseudo\-spin field ${\bf m}({\bf r},\tau) = (\sin \theta \cos \phi,
\sin \theta \sin \phi, \cos \theta)$, in the presence of both
tunneling and gates, is given by~\cite{doublelayerrefs}
\begin{eqnarray}
   S^E[{\bf m}] &=& \int \! d\tau \, d^2r \,
   \left\{
      \frac{-i\dot{{\bf m}} \cdot {\bf A}}{4 \pi \ell^2} +
      \frac{\rho}{2}
      \left[
         \left( \nabla m_x \right)^2 + \left( \nabla m_y
         \right)^2
      \right]
   \right. \nonumber \\ && \mbox{} +
   \left.
      \beta_0 m_z^2 -
      \frac{1}{2 \pi \ell^2}
         \left[ t m_x - V(x) m_z \right]
   \right\}.
\label{eq-full}
\end{eqnarray}
The first term is the Berry phase, conveniently expressed as
$\dot{{\bf m}} \cdot {\bf A} = \dot{\phi} \, (1 - \cos \theta)$.  The
gradients are the exchange terms, with $\rho$ the pseudospin
stiffness.  The term involving $\beta_0$ is a hard-axis anisotropy,
and in the following term, $t$ is the tunneling amplitude which acts
as an in-plane pseudofield.  Finally, $V(x)$ is the gate potential
which can be adjusted {\it in situ} for appropriately fabricated
double-layer samples. The local filling factor for each layer is
$\nu_1 = (1 + \cos \theta)/2$, $\nu_2 = (1 - \cos \theta)/2$, with the
total filling factor $\nu = \nu_1 + \nu_2 = 1$.  The parameters $\rho$
and $t$ in Eq.~(\ref{eq-full}) depend on $m_z$. We require their
expansion to quadratic order in $\vartheta = \theta - \pi/2$,
$\rho(m_z) = \rho_0 + \rho_1 \vartheta^2$ and $t(m_z) = t_0 + t_1
\vartheta^2$, and shall use the Hartree-Fock results
\cite{bettercoming} $\rho_1 = -\rho_0$ and $t_1 = -t_0/2$.

We limit our attention to finite-width ($w$) systems for which the
soliton physics is particularly simple.  The flexural bending
mode~\cite{winter62,braun96} of a soliton in the $x$ direction of a
two-dimensional system has a spectrum given by $\epsilon(k_y) = 4 \pi
\ell^2 k_y \sqrt{\rho_0 (2 \beta_0 + \rho_0 k_y^2)}$, where $k_y$ is
the transverse wavevector.  The finite-size gap of these modes can be
estimated by setting $k_y = \pi /w$. For a given temperature $T$, the
quasi-one-dimensional limit results if flexural modes are frozen out,
{\it i.e.}, if the transverse sample size is smaller
than~\cite{braun96} $w(T) = (\pi / \tau) [(\rho_0 / \beta_0)(1 +
\sqrt{1 + \tau^2})]^{1/2}$, where $\tau = k_B T / (4 \pi \ell^2
\beta_0)$. For the sample parameters we give below, $w(1\,{\rm K}) =
718\,\text{nm}$, whereas $w(100\,{\rm mK}) = 7\,\mu\text{m}$. Assuming
the transverse sample size to be less than this value, we can
trivially integrate over the $y$ direction.

The estimates for the tunneling rates given below are based on the
following \cite{hanna98} parameter values: $2 \pi \ell^2 \beta_0
\approx 2.3 \, \text{meV}$, $\rho_0 \approx 0.024\,$meV, $t_0 \approx
0.1\,$meV, and $\ell = 11.8\,$nm.  We limit our attention to the
regime $V(x) \ll 2 \pi \ell^2 \beta_0$ and look for solutions to the
equations of motion with $\theta = \pi / 2$.  For $V(x) = 0$, the
solution $\phi_0 (x - X) = 4 \arctan e^{\pm (x - X) / \delta}$
satisfies the equation of motion, where $\delta = \sqrt{2 \pi \rho_0 /
  t_0} \, \ell$. The solution $\phi_0$ describes a static $2 \pi$
soliton with width $\delta$ centered at position $X$. The pseudospins
rotate in the $xy$ plane by $2 \pi$ as we move through the soliton,
beginning and ending at $\phi_0 (\pm \infty) = 0, \, 2 \pi$.  Quantum
effects are incorporated by expanding about this classical solution,
and dynamics is described through a canonical transformation to
collective coordinates as we discuss below \cite{braun96}.  We include
the effects of the gates self-consistently by substituting the soliton
solution $(\theta,\phi) = (\pi/2, \phi_0)$ into the full action.  This
is valid so long as the soliton energy $E_{\text{soliton}} = 8 L_y
\rho_0 / \delta$ is much larger than the effective potential.

Expanding to quadratic order in $\vartheta (x,\tau) = \theta(x,\tau) -
\pi/2$, and performing the functional integral over $\vartheta$ (in
the partition function), we obtain
\begin{eqnarray}
   \frac{S_{\text{eff}}}{L_y } &=& \! \int \!\! d\tau dx \!
   \left[
      \frac{-i}{4 \pi \ell^2} \dot{\phi}  +
      \frac{\rho_0}{2}
      \left(
         \left( \partial\phi \right)^2 +
         \frac{1}{c^2} \dot{\phi}^2
      \right) -
      \frac{t_0 \cos \phi}{2 \pi \ell^2}
   \right. \nonumber
\\ && \mbox{}  -
   \left. \frac{V^2(x)}{a^4 \beta_0^2}
      \left(
         \rho_0 \left(\partial\phi \right)^2 -
         \frac{t_0 \cos \phi}{2 \pi \ell^2}
      \right)
   \right],
   \label{eq-s.quad}
\end{eqnarray}
where $c = 2 \pi \ell^2 \sqrt{8 \beta_0 \rho_0}$ is the spin-wave
velocity.  This equation is valid for $t_0 / (2 \pi \ell^2 \beta_0)
\ll 1$ and $4 \rho_0 / (\delta^2 \beta_0) \ll 1$.  For the parameter
values given above, $t_0 / (2 \pi \ell^2 \beta_0) \approx 0.04$ and $4
\rho_0 / (\delta^2 \beta_0) \approx 0.17$.

To describe motion of the soliton, we perform a canonical
transformation to collective coordinates.  This entails elevating $X$
to a dynamical variable~\cite{rajaraman87,braun96}, $X \rightarrow
X(\tau)$, and introducing a constraint to preserve the degrees of
freedom.  We thus write $\phi (x,\tau) = \phi_0 (x - X(\tau)) +
\varphi(x - X(\tau), \tau)$ and expand to quadratic order in
$\varphi$, where $\phi_0$ now describes a moving soliton and $\varphi$
describes a dissipative spin-wave field, which is required to be
orthogonal to the zero mode of the soliton field. We incorporate this
constraint, explicitly given by $\int\! dx\, \phi'_0(x) \varphi
(x,\tau) = 0$, via the Fadeev-Popov technique. Briefly, the
procedure~\cite{braun96} is to insert into the functional integral the
identity $\int\! {\cal D}X\, \delta (Q) \det (\delta Q/\delta X)
\equiv 1$, where $Q[X] = \int\!  dx\, \phi'_0(x-X) \phi (x,\tau)$.  We
then expand to second order in both $\varphi$ and $\dot{X}/c$.  For
the bounce solutions considered below, $|\dot{X}/c| \le 0.26$.  This
yields a description of the solitons, the spin waves, and the dynamic
(non-linear) coupling between them.

Integrating over the spin waves, and including the gate potential,
yields an action corresponding to a particle (soliton) at position
$X$, with bare mass $M$, moving in a potential
$\widetilde{V}_{\text{eff}} (X)$, with dissipation described by a
nonlocal kernel $K(\tau)$:
\begin{eqnarray}
   S_{\text{eff}} [X] &=& \int_{-\beta/2}^{\beta/2} \! d\tau\,
   \left\{ \frac{1}{2} M \dot{X}^2 +
   \widetilde{V}_{\text{eff}} (X) \right. \nonumber \\
   &+& \left. \frac{1}{2}
   \int_{0}^{\tau} \! d\tau'\, K(\tau - \tau')
   \left[ X(\tau) - X(\tau') \right]^2 \right\}.
   \label{eq-SX}
\end{eqnarray}
For low temperatures ($k_B T \ll \hbar \omega_0 = 2 \hbar c/\delta$),
the kernel is given by $K(\tau) = (1/\pi) \int_0^\infty \! d\omega\,
J(\omega) e^{-\omega |\tau|}$.  The spectral density $J(\omega) =
(\omega/\omega_0 \delta^2) \sqrt{\omega^2 - \omega_0^2}\,
\Theta(\omega - \omega_0)$ vanishes for $\omega < \omega_0$ ($\Theta$
is the step function), and the kernel $K(\tau)$ is exponentially
suppressed for $|\tau| >1/ \omega_0$.  In the parameter range we are
interested in, $k_B T \ll \hbar \omega_b \ll \hbar \omega_0$
($\omega_b$ is the bounce frequency---see below), we can expand
$X(\tau) - X(\tau') \approx (\tau - \tau') \dot{X}(\tau')$, and the
effect of damping reduces to a renormalization of the soliton mass
$M$.  This additive renormalization is intensive with respect to the
transverse length $L_y$ of the sample~\cite{int.note}, whereas the
bare mass is extensive (proportional to $L_y$), and so we shall not
consider this additive piece in what follows.  For the estimates we
give below, $\omega_b/\omega_0 \le 0.11$, and $k_B T_C/ \hbar \omega_b
= 0.47,\ 0.16,\ 0.13,\ {\rm and}\ 0.13$ respectively for the entries
listed in Table~\ref{tab-tunnel}, where $T_C$ is the crossover
temperature (see below).

To leading order, the bare mass is given by $M =\hbar^2 L_y / (4 \pi^2
\ell^4 \beta_0 \delta)$, and the effective potential by
\begin{equation}
   \widetilde{V}_{\text{eff}} (X) =
   \frac{-3 L_y \rho_0}{8 (2 \pi \ell^2)^2 \beta_0^2}
   \int_{-\infty}^{\infty} dx \, V^2(x)
   \Bigl[ \partial\phi_0 \left( x - X(\tau) \right) \Bigr]^2.
\end{equation}
This effective potential can be understood in terms of the change in
the classical soliton energy density when the $m_z =0$ values for $t$
and $\rho$ in Eq.~(\ref{eq-full}) are replaced by their values at $m_z
= - V(x) /4 \pi \ell^2 \beta_0$, {\it i.e.}, when the pseudospin
polarization produced by the bias potential is accounted for. In
Eq.~(\ref{eq-SX}), we have neglected phase terms \cite{braun96}. All
such terms vanish for {\em incoherent} tunneling, as in the case of
tunneling out of a metastable state considered below.

To obtain an explicit expression for $\widetilde{V}_{\text{eff}}$, we
should specify the form of the gate potential $V(x)$.  We choose two
simple square wells.  One applies a gate voltage of $-V_1$ over a
width $w_1$, $-V_2$ over a width $w_2$, and zero voltage elsewhere.
The inside edges of the gates are a distance $2 x_0$ apart. This
yields an effective potential explicitly given by
\begin{equation}
   \widetilde{V}_{\text{eff}} (X) = v \sum_{i=1}^2 (-1)^{i}
   V_i^2
   \left[
       \tanh \bar{x}_i - \tanh (\bar{x}_i - \bar{w}_i)
   \right],
\end{equation}
where $v = -3L_y\rho_0/[2(2 \pi \ell^2)^2\delta \beta_0^2]$,
$\bar{x}_i = [X - (-1)^i x_0] / \delta$, and $\bar{w}_i = (-1)^i w_i /
\delta$. Self-consistency demands that $E_{\text{soliton}} \gg
\widetilde{V}_{\text{eff}}$, which places an upper bound on the gate
voltage of $V_i^{\text{max}} \approx 2.3\,\text{meV}$.

We will consider tunneling out of metastable states.  Once a soliton
tunnels out, it does not return.  We therefore put $|V_1| < |V_2|$ and
$w_1 \ll w_2$.  The potential is schematically shown in
Fig.~\ref{fig-pot}, with the coordinates shifted as outlined below.
The objective is to calculate the tunneling rate out of this
metastable state.

\begin{figure}
   \begin{center}
     \epsfxsize=8.5cm \epsffile{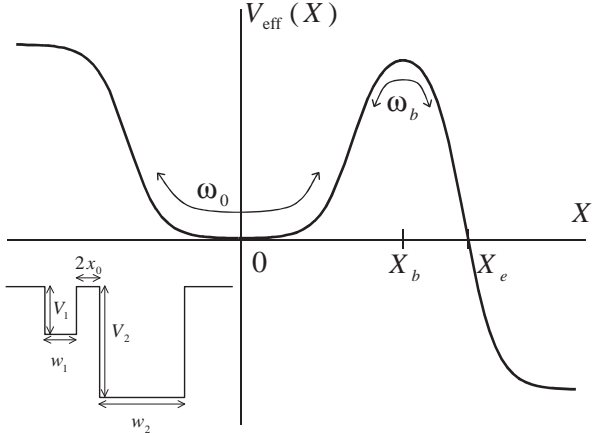}
   \end{center}
   \caption{Effective potential produced by the gate voltages in
     the ``metastable'' configuration.  The inset shows the ``bare''
     gate potential.}
   \label{fig-pot}
\end{figure}

It is helpful to shift coordinates so that the metastable minimum is
now at the origin and $V_{\text{eff}}(0) = 0$ (See
Fig.~\ref{fig-pot}).  There is a frequency $\omega_0 =
\sqrt{V''_{\text{eff}}(0) / M}$ associated with the curvature of the
metastable minimum, and a frequency $\omega_b =
\sqrt{V''_{\text{eff}}(X_b) / M}$ associated with the barrier.  The
tunneling rate is given by $\Gamma = K \omega_0 \left( S_0 / 2 \pi
\right)^{1/2} e^{-S_0}$, where $K$ is a constant whose calculation
requires explicit knowledge of the bounce trajectory. The exponent
$S_0$ is the action evaluated along the minimal (bounce) trajectory,
which goes from $X=0$ to $X=X_e$ and back to $X=0$, while $\tau$ goes
from $-\beta/2$ to $+\beta/2$, with $\beta \rightarrow \infty$.  The
crossover temperature $T_C$ defines the boundary between
thermally-dominated transitions ($T > T_C$) and tunneling-dominated
transitions ($T < T_C$).  Both transitions show exponential behavior
and one usually assumes that the exponent dominates the prefactor,
leading to the expression $k_B T_C = V_{\rm eff} (X_b) / S_0$.

The tunneling rate, along with the results for the crossover
temperature $T_C$, the classical action $S_0$, and the attempt
frequency $\omega_0$ are listed in Table~\ref{tab-tunnel} for various
gate voltage profiles and transverse sample sizes. These entries were
evaluated using the typical model parameters quoted earlier which
yield a soliton width of $\delta = 14.5$\,nm and a bare soliton mass
of $M \approx m^{\ast} (0.44 L_y / \ell) $, where $m^{\ast}$ is the
conduction band effective electron mass in GaAs. We conclude from
Table~\ref{tab-tunnel} that voltage profiles for which quantum
tunneling will dominate at accessible temperatures are achievable with
current submicron lithographic technology.

\begin{table}
\squeezetable
\caption{Tunneling rate $\Gamma$, crossover temperature $T_C$,
   and attempt frequency $\omega_0$ for several gate widths
   $w_1$, gate separations $2 x_0$, and sample sizes $L_y$.
   All entries have
   $w_2 = 400$\,nm, $V_1 = -0.75$\,meV, and $V_2 = -1.00$\,meV,
   except for the final entry, which also has $V_1 = -1.00$\,meV.}
\begin{center}
\[\begin{array}{ccccccc}
   \hline \hline
   L_y & w_1 & 2x_0 & \omega_0 & S_0 & \Gamma & T_C \\
   \mbox{[$\mu\text{m}$]} & [\text{nm}] & [\text{nm}] & [\text{GHz}] &
   &  & [\text{mK}]\\ \hline
   0.3 & 100 & 100 & 36 & 12.0 &
      93\,\text{kHz} & 150
      \\
   0.6 & 100 &  50 & 36 & 12.5 &
      66\,\text{kHz} &  267
      \\
   1.0 &  50 &  25 & 187 & 8.9 &
      63\,\text{MHz} &  367
      \\
   1.0 &  20 &  20 & 448 & 5.3 &
      19\,\text{GHz} &  463
      \\ \hline \hline
\end{array}\]
\end{center}
\label{tab-tunnel}
\end{table}

There are at least three properties of pseudospin solitons which
should make their existence and their motion experimentally
observable: moving solitons disturb the charge balance between the
layers; there exists a local electrical current circulating about the
axis of the soliton; and the local quasiparticle gap is suppressed in
the vicinity of the soliton.  These local changes can be detected
respectively by electrostatic, magnetic, and tunneling probes of the
2DES.  We discuss each briefly.

That moving solitons transfer charge between layers may be seen by
revisiting the classical equations of motion for the action in
Eq.~(\ref{eq-full}) and trying to find {\em moving} soliton solutions.
Proceeding in powers of $v/c$, where $v$ is the soliton velocity and
$c$ the spin wave velocity in Eq.~(\ref{eq-s.quad}), we write
$\theta_{\text{cl}} (s) = \pi/2 + (v/c) \theta_1 (s) + O(v^2/c^2)$ and
$\phi_{\text{cl}} (s) = \phi_0(s) + (v/c) \phi_1 (s) + O(v^2/c^2)$,
where $\phi_0 (s)$ is the previous solution for the {\em static}
soliton, and $s = x - v \tau$ is the coordinate of the moving frame.
Note that this ansatz satisfies the equations of motion {\em exactly}
for $v=0$.  To linear order in $v/c$, we find that $\phi_1(s) = 0$,
while
\begin{equation}
   \theta_1(s) = \frac{\pm (c/2\delta) \,
   \text{sech} \left[ (s - s_0) / \delta \right]}
   {(a^2 \beta_0 + t_0) -
   6 t_0 \text{sech}^2 \left[ (s - s_0) / \delta \right]}.
\label{eq-move}
\end{equation}
The local charge transfer produced by the passing soliton is related
to the charge transfer produced by the static bias potential which, by
virtue of the difference in layer energies it produces, gives the same
rate of change of the interlayer phase $\phi$. This result shows that
whereas a static soliton pseudospin rotates entirely in the $xy$
plane, a {\em moving soliton pseudospin rises out of the plane} by an
amount proportional to the velocity as it undergoes its $2 \pi$
rotation. This is analogous to closely related models of anisotropic
ferromagnets where exact moving soliton solutions are available
\cite{braun96a}.  In physical terms, a moving soliton transfers charge
between the layers, altering the electric potential profile outside of
the sample, which can be measured using a scanning single-electron
transistor\cite{yacoby}. The total transferred charge for a soliton is
given by $Q_e = L_y n e \int \!dx\, \cos \theta_{\text{cl}} \approx
\mp \pi e n L_y (\dot{X} / c) \sqrt{2 \rho_0 / \beta}$, with
corrections of order $(\dot{X} / c)^2$ and $t_0 / (a^2 \beta_0)$, and
where $n \approx 10^{11} \text{cm}^{-2}$ is the electron density.
Using the parameter values given above, a system size of $L_y = 600\,
\text{nm}$, and taking for $\dot{X}$ the maximal velocity along the
bounce trajectory, we obtain $|Q_e| \approx 2e$. This value can be
increased either by inducing a larger soliton velocity, or by going to
a larger sample. These charges should appear, for example, as a
soliton tunnels from a metastable state.

Currents, much like those near a vortex in a superconductor, circulate
around the axis formed by a soliton line and produce a magnetic field
which could in principle be measured.  The microscopic operator for
the current flowing between layers is the time derivative of the
layer-polarization operator in the Heisenberg representation.  The
only contribution to the commutator comes from the interlayer
tunneling term in the microscopic Hamiltonian, and is proportional to
the operator for the $\hat y$ component of the pseudospin.  Taking its
expectation value we find for the 3D current between the layers (with
the bottom to top layer direction taken as positive) $j_{BT}(x) = - t
\sin (\phi_0 (x))/ \hbar 2 \pi \ell^2 $.  Currents flow in opposite
directions on opposite sides of the soliton.  The currents flowing
within the bottom (B) and top (T) 2D layers are oppositely directed
and given by \cite{doublelayerrefs} $j_B(x) = - j_T(x) = \rho \partial
\phi_0/\hbar$. For a stationary soliton these currents satisfy the
local charge conservation condition $ d j_B(x)/dx = - d j_T(x)/dx =
j_{BT}(x)$. The current per unit length flowing around the soliton is
$\sim e \rho / \hbar \delta$ and the characteristic loop area of the
circulating currents is $\sim d \delta$ where $d$ is the layer
separation.  For typical parameters, these currents produce a magnetic
induction directed along the soliton axis with magnitude $\sim 0.1
\mu_B / \ell^2 \delta$, which could in principle be detected by a
SQUID or with cantilever-based\cite{harris} magnetometers.

Microscopic calculations can also be used to evaluate the local gap
for charged excitations as a function of position near a stationary
soliton.  We find\cite{unpub} that the local gap is always reduced
near the soliton center and is given by $E_{\rm gap}(x) = E_{\rm
  gap}(\infty) - 4 t(1 - \cos(\phi_0(x))$. This change in spectral
properties might be visible in scanning tunneling probes of 2DES's.
This gap suppression is local and will increase the local thermally
activated quasiparticle resistivity $\rho_{xx}$ at the soliton center
by a factor of $ \exp (8 t /k_B T) \sim \exp( 10/ T[{\rm Kelvin}])$
which can easily be very large.  When a soliton tunnels away from a
region defined by a set of voltage probes, we can expect a large
signal to be seen in the conductances measured by those probes.

In all of the above, we have assumed the external magnetic field was
oriented perpendicular to the layers.  But this need not be the case,
and it may in fact be preferable to add an in-plane component to the
field.  In the presence of an in-plane field ${\bf B}_\parallel = (0,
-B_\parallel, 0)$, the energy~\cite{hanna98} of the soliton is lowered
and is given by $E_{\text{sol}} = L_y \rho_0 (8/\delta - 4 \pi^2 d
B_\parallel / \Phi_0)$, where $\Phi_0$ is the flux quantum. The
benefit, in the present context, of an in-plane field is that it may
be used to thermally create solitons at arbitrarily low-temperatures
and thus ensure that solitons exist in the sample.

In summary, we have shown that pseudospin solitons may be pinned and
manipulated by applying external gates to the system.  One can extend
the analysis to the case of a double-well potential, which should
exhibit an externally tunable tunnel splitting, and to the case of a
periodic potential. The physics in the latter case should be driven by
Berry and topological phases, and should be strongly affected---and
possibly controlled---by the gate voltages. This opens the door to a
variety of Aharonov-Bohm-type investigations, as well as soliton
delocalization and band formation \cite{braun96}.  We have also shown
that solitons have an electrical current circulating about their axis,
that the local quasiparticle gap is suppressed in the vicinity of the
soliton, and also that soliton motion induces a local charge imbalance
between the layers.  These properties may provide a means through
which the soliton dynamics can be experimentally probed.

This work has been supported by the Swiss NSF, and by the US NSF (DMR
97-14055, PHY 94-07194).






\title{Erratum: Quantum Dynamics of Pseudospin Solitons in
  Double-Layer Quantum Hall Systems [Phys.\ Rev.\ Lett.\ {\bf 83},
  1411 (1999)]}
\author{Jordan Kyriakidis, Daniel Loss, and A. H. MacDonald}

\maketitle
\widetext

\vspace{.5cm}

The expressions, correct to Hartree-Fock order, for the pseudospin
stiffness $\rho$ and tunneling amplitude $t$ are given by
$\rho=\rho_0$ and $t=t_0$.  Consequently, the right-hand side of
Eqs.~(4) and (5) of our Letter should be multiplied by $1/2$.  Some of
the entries in Table~I are also affected and more representative
values are given below.  The conclusions and all other results remain
unaffected.

\begin{table}[h]
\squeezetable
\caption{Tunneling rate $\Gamma$, crossover temperature $T_C$,
   and attempt frequency $\omega_0$ for several gate widths
   $w_1$, gate separations $2 x_0$, and sample sizes $L_y$,
   assuming negligible dissipation (see text).  All entries have
   $w_2 = 400$\,nm, $V_1 = -0.75$\,meV, and $V_2 = -1.00$\,meV,
   except for the final entry, which also has $V_1 = -1.00$\,meV
   (and the same $V_2$ and $w_2$ as the other entries).}
\[\begin{array}{ccccccc}
   \hline \hline
   L_y & w_1 & 2x_0 & \omega_0 & S_0 & \Gamma & T_C \\
   \mbox{[$\mu\text{m}$]} & [\text{nm}] & [\text{nm}] & [\text{GHz}] &
   & [\text{KHz}] & [\text{mK}]\\ \hline
   0.6 & 100 & 100 &  26 & 17.0 &   0.5 & 106 \\
   1.0 & 100 &  50 &  25 & 14.7 &   5   & 188 \\
   2.5 &  50 &  25 & 132 & 15.8 &  63   & 260 \\
   2.5 &  20 &  20 & 317 &  9.4 & 302   & 327 \\
   \hline \hline
\end{array}\]
\end{table}


\begin{thebibliography}{10}
\bibitem{bettercoming} Microscopic evaluation of quantum fluctuation
  corrections to the Hartree-Fock results for $\rho (m_z)$ and
  $t(m_z)$ are feasible and would be useful.  Once available, these
  corrections can be used to adjust some of the quantitative estimates
  made below.
\bibitem{int.note} The intensive renormalization arises from the
  expansion of both $\dot{X}$ and $\varphi$.  The first extensive
  contribution to $S_{\text{eff}}$ from the dissipation is at least
  fourth order in $\dot{X}$.  For details, see Eqs.~(B1) through (B4)
  of Appendix~B in Ref.~\cite{braun96}.
\end{thebibliography}
\end{document}